\documentclass{article}
\usepackage[utf8]{inputenc}

\newsavebox{\foobox}
\newcommand{\slantbox}[2][0]{\mbox{%
        \sbox{\foobox}{#2}%
        \hskip\wd\foobox
        \pdfsave
        \pdfsetmatrix{1 0 #1 1}%
        \llap{\usebox{\foobox}}%
        \pdfrestore
}}
\newcommand\unslant[2][-.25]{\slantbox[#1]{$#2$}}

\newcommand{\mdelta}{\text{\unslant[-.18]\delta}}

\usepackage[left=2.5cm,right=2.5cm, top=1in, bottom=1in]{geometry}

\usepackage{isomath}
\usepackage{euscript}
\usepackage{amssymb}
\usepackage{amsfonts}
\usepackage{amsmath}
\usepackage{amsbsy}
\usepackage{epsfig}
\usepackage{amsthm}
\usepackage{amscd}
\usepackage{amstext}
\usepackage{verbatim}
\usepackage{amsmath}
\usepackage{cancel}
\usepackage{capt-of}
\usepackage{empheq}
\usepackage{slashed}
\usepackage{xcolor}
\usepackage{setspace}

\usepackage[colorlinks=true, urlcolor=violet, linkcolor=blue, citecolor=red, hyperindex=true, linktocpage=true]{hyperref}

\def\ben{\begin{equation}}
\def\een{\end{equation}}

\let\a=\alpha   \let\d=\delta 
   \let\k=\kappa
     
\let\s=\sigma

\let\pa=\partial
\def\be{\begin{equation}}
\def\ee{\end{equation}}
\def\beq{\begin{equation}}
\def\eeq{\end{equation}}
\def\ba{\begin{array}}
\def\ea{\end{array}}

\def\dalemb#1#2{{\vbox{\hrule height .#2pt
       \hbox{\vrule width.#2pt height#1pt \kern#1pt
               \vrule width.#2pt}
       \hrule height.#2pt}}}

\newcommand{\bea}{\begin{eqnarray}}
\newcommand{\eea}{\end{eqnarray}}

\def\ocal{{\mathcal{O}}}

  \usepackage{titlesec}

\renewcommand\theparagraph{\thesubsubsection.\arabic{paragraph}}

\makeatletter

\DeclareRobustCommand\bfseriesitshape{%
  \not@math@alphabet\itshapebfseries\relax
  \fontseries\bfdefault
  \fontshape\itdefault
  \selectfont
}

\titleformat{\paragraph}
{\normalfont\normalsize\bfseries}{\theparagraph}{1em}{}
\titlespacing*{\paragraph}
{0pt}{3.25ex plus 1ex minus .2ex}{1.5ex plus .2ex}
\def\toclevel@paragraph{4}
\def\l@paragraph{\@dottedtocline{4}{7em}{4em}}
\makeatother

\begin{document}

\allowdisplaybreaks

\thispagestyle{empty}

\begin{center}

{ \Large {\bf
Resistivity bound for hydrodynamic bad metals
}}

\vspace{1cm}

Andrew Lucas and Sean A. Hartnoll

\vspace{1cm}

{\small{\it 
 Department of Physics, Stanford University, \\
Stanford, CA 94305-4060, USA \\
\vspace{0.3cm}

 }}

\vspace{1.6cm}

\end{center}

\begin{abstract}
We obtain a rigorous upper bound on the resistivity $\rho$ of an electron fluid whose electronic mean free path is short compared to the scale of spatial inhomogeneities. When such a hydrodynamic electron fluid supports a non-thermal diffusion process -- such as an imbalance mode between different bands  -- we show that the resistivity bound becomes $\rho \lesssim A \, \Gamma$. The coefficient $A$ is independent of temperature and inhomogeneity lengthscale, and $\Gamma$ is a microscopic momentum-preserving scattering rate. In this way we obtain a unified and novel mechanism -- without umklapp -- for $\rho \sim T^2$ in a Fermi liquid and the crossover to $\rho \sim T$ in quantum critical regimes. This behavior is widely observed in transition metal oxides, organic metals, pnictides and heavy fermion compounds and has presented a longstanding challenge to transport theory. Our hydrodynamic bound allows phonon contributions to diffusion constants, including thermal diffusion, to directly affect the electrical resistivity.

\end{abstract}

\vspace{7cm}

\noindent \texttt{ajlucas@stanford.edu}
\\
\texttt{hartnoll@stanford.edu}

\pagebreak
\setcounter{page}{1}

\tableofcontents

\section{Introduction}

\subsection{Motivation: Scattering versus resistivity in strange metals}

Electrical resistance arises due to microscopic scattering of charge carriers. The resistivity, however, is a macroscopic observable sensitive to the rate of momentum relaxing scattering events. The challenge of unconventional or `strange' metals is largely associated with relating momentum-conserving to momentum-relaxing scattering rates.  
Most notably, a $T$-linear resistivity is commonly observed close to quantum critical points in strange metals \cite{Sachdev:2011cs}. And indeed, a $k_{\mathrm{B}} T/\hbar$ scattering rate is both characteristic of quantum criticality \cite{subir} and widely seen in strange metals in single particle \cite{ARPES1} as well as transport \cite{OPTICAL1,andy} observables. Nonetheless, because the scattering of electrons by quantum critical fluctuations is typically momentum-preserving, this rate does not directly control the resistivity in theoretical models of such systems. The interactions between electrons and collective quantum critical fluctuations are often strong, and hence any momentum transferred from charge carriers to quantum critical modes is rapidly returned \cite{Mahajan:2013cja, Hartnoll:2014gba, Patel:2014jfa}. For this reason, among others, a compelling microscopic theory of $T$-linear quantum critical resistivity has remained elusive.

Away from the $T$-linear resistivity characterizing `quantum critical fans', a $T^2$ resistivity is widely observed in strongly correlated metals including transition metal oxides, pnictides, organic metals and heavy fermion compounds \cite{KADOWAKI1986507, Jacko2009}. The $T^2$ scaling is usually not considered `strange' because electronic umklapp scattering is a conventional mechanism that can lead to such a resistivity. However, that may be a premature conclusion.
In this paper we will describe a new scenario in which a momentum-preserving microscopic scattering rate directly controls the resistivity $\rho$. This will lead to an alternate path to $\rho\propto T^2$, without umklapp, that transitions to $\rho\propto T$ when the momentum-conserving scattering rate changes from $T^2$ to $T$. In our theory, the mechanism for momentum relaxation -- charge density inhomogeneities on length scales longer than the electronic mean free path --   does not change across the phase diagram.

\subsection{Hydrodynamic electron fluids}

Denote by $\xi$ the characteristic length scale of spatial inhomogeneities, and by  $\ell_\text{ee}$ the mean free path of electrons due to momentum-preserving interactions (either with electrons or phonons).  When the interactions are strong enough that $\xi \gg \ell_\text{ee}$, then the electron fluid will be locally equilibrated with a well-defined local momentum density that is only relaxed on the length scale $\xi$. In such cases the conductivity can be calculated entirely within hydrodynamics \cite{KS06, KS11, Lucas:2015lna, PhysRevB.93.075426}, which is the theory of macroscopic, long-wavelength perturbations from equilibrium. Bad metals in particular have extremely short mean free paths \cite{PhysRevLett.74.3253,
RevModPhys.75.1085, MIR}. The same strong interactions that cause the absence of quasiparticles in these materials also lead to efficient local thermalization. 


The main technical result in this paper is that the electrical resistivity of a hydrodynamic electron fluid is non-perturbatively (in disorder amplitude) bounded above by a certain spatial average of a matrix of `incoherent' diffusion constants. These incoherent diffusive processes decouple from the conserved momentum density and hence from `momentum drag' effects \cite{Hartnoll:2014lpa}, and are proportional to the microscopic mean free path $\ell_{\mathrm{ee}}$. We prove our bound using a variational principle for hydrodynamic entropy production developed in \cite{Lucas:2015lna}.  The most well-known incoherent diffusion mode is thermal diffusion \cite{KS11}. However, at low temperatures thermal diffusion is suppressed. A central observation in this paper is that the presence of an additional, non-thermal diffusive mode has very significant and interesting consequences for low temperature transport in hydrodynamic metals. An example of such a mode is an `imbalance' mode resulting from multiple Fermi bands or pockets, each with an approximately independently conserved number density. See figure \ref{fig:Imb} below. With such a non-thermal diffusive mode we show that, if our bound is saturated, the temperature dependence of the resistivity is directly determined by a microscopic electronic scattering rate. In this way we obtain a unified description of $T^2$ and $T$-linear resistivity in hydrodynamic electron fluids.

With the inhomogeneity fixed, the microscopic scattering rate can increase arbitrarily at high temperatures. The hydrodynamic fluid description only improves in this limit as the mean free path gets shorter. The framework we have introduced therefore lends itself to the study of bad metals.

\section{Resistivity bound in hydrodynamics}

\subsection{Hydrodynamic equations}
\label{sec:hydroequations}

Consider an electron fluid with $N$ conserved scalar quantities, with associated densities $n^A$.  Denoting the conserved currents of these scalars by $j^A$, we find $N$ conservation laws: $\dot n^A + \nabla \cdot j^A = 0 \,.$ For a minimal linearized hydrodynamics the conserved densities would be the charge density $n$ and entropy density $s$, and the corresponding currents would be the electric and entropy currents, $j$ and $j^{\mathrm{Q}}/T$.
We are interested in the case where there are additional conserved quantities, such as the imbalance modes discussed in the introduction.   There is also a conserved momentum density $\pi$, and a corresponding conservation law:   $\dot \pi_i + \pa_j \tau_{ji} = 0 \,.$
Here $\tau$ is the stress tensor.

To complete the hydrodynamic description, one must write down the constitutive relations that give the currents in terms of a gradient expansion of the densities (or their conjugate sources). For the currents $j^A$ the constitutive relations are
\be\label{eq:consgen}
j^A = n^A \, v  - \Sigma_0^{AB} \nabla \mu_B + \cdots \,.
\ee
Here $\mu_A$ is a  `chemical potential' conjugate to the conserved density $n^A$.  For instance, the source conjugate to the charge density is the chemical potential $\mu$, the entropy density $s$ is conjugate to the temperature $T$, and the momentum density $\pi$ is conjugate to the velocity $v$.  Within linear response, the densities and sources are related by a matrix $\chi$ of thermodynamic susceptibilities: 
$\delta n^A = \chi^{AB} \delta \mu_B$, and $\pi^i = {\mathcal M} \, v^i \,.$
In Galilean-invariant cases, the thermodynamic susceptibility ${\mathcal M}$ equals the mass density.
The first term in the constitutive relations (\ref{eq:consgen}) describes how the momentum drags the scalar densities, while  $\Sigma_0^{AB} = \Sigma_0^{BA}$ are the `incoherent' transport coefficients. The constitutive relation for the stress tensor is
\be\label{eq:conspi}
\tau_{ij}  =  \delta_{ij} P - \zeta \delta_{ij} \nabla \cdot v - \eta \left(\pa_i v_j + \pa_j v_i - \frac{2}{d} \delta_{ij} \nabla \cdot v \right) + \cdots \,.
\ee
Here $P$ is the pressure and $\zeta$ and $\eta$ are the bulk and shear viscosities, respectively.   Note that \begin{equation}
\mathrm{d}P =  n^A \mathrm{d}\mu_A.  \label{eq:dP}
\end{equation}

Momentum relaxation is incorporated into the above equations by allowing all of the coefficients, thermodynamic susceptibilities and equilibrium expectation values and sources to be space-dependent. This spatial dependence must be sufficiently long wavelength that it is consistent with the hydrodynamic derivative expansion in the constitutive relations.

\subsection{Variational principle and resistivity bound}
\label{sec:varprin}

We will establish an upper bound on the resistivity using a hydrodynamic variational principle, building upon the results in \cite{Lucas:2015lna}.  The basis for this technique is that the entropy production is smallest on the solutions to the equations of motion.   In this regard the variational principle is similar in spirit to that for the kinetic equation \cite{zimanshort,ziman}, Thomson's principle for resistor networks \cite{levin}, and holographic conductivity bounds \cite{Grozdanov:2015qia, Grozdanov:2015djs}.
The rate of entropy density production in linearized hydrodynamics is quadratic in the hydrodynamic variables \cite{Hartnoll:2016apf}, and can be expressed in terms of the currents $j^A$ and velocities $v$ as
\be\label{eq:sdot}
T \dot s = \eta \left(\partial_i v_j + \partial_j v_i - \frac{2}{d} \mdelta_{ij} \partial_k v_k\right)^2 + \zeta (\nabla \cdot v)^2 + \left(j^A - n^A v\right) (\Sigma_0^{-1})_{AB}\left(j^B - n^B v\right) \,.
\ee
Now define the functional
\be\label{eq:functional}
\mathcal{R}[v,j^A]  \equiv \frac{ \displaystyle \int \frac{\mathrm{d}^dx}{V} \, T \dot s }{\left(\displaystyle \int \frac{\mathrm{d}^dx}{V} \, j_x \right)^2} \,,
\ee
where the entropy production in the numerator is given by (\ref{eq:sdot}), viewed as a function of $v$ and $j^A$, and in the denominator $j_x$ is a component of the electric current. $V$ is the volume.  Following \cite{Lucas:2015lna}, we prove in Appendix \ref{sec:proof} that (\emph{i}) if the functional (\ref{eq:functional}) is varied over velocities $v$ and currents satisfying $\nabla \cdot j^A = 0$, then it is minimized on stationary solutions to the hydrodynamic equations of motion in the presence of a uniform applied electric field $E_x$ and (\emph{ii}) the functional evaluated on these solutions is equal to the electrical resistivity $\rho_{xx} = 1/\sigma_{xx}$, where in the presence of the applied electric field: $j_x = \sigma_{xx} E_x$.  In this last equation, we do not impose any constraints on the average currents $\int \mathrm{d}^dx j^A$; instead, we demand that the electric field is the only source. It follows that the resistivity is bounded by
\be\label{eq:var}
\rho_{xx} = \min_{\nabla \cdot j^A = 0} \mathcal{R}[v,j^A] \; \leq \; \mathcal{R}[v,j^A] \Big|_{\nabla \cdot j^A = 0}\,.
\ee

With translation invariance, none of $\eta,\zeta,\Sigma_0^{-1}$ or $n$ in (\ref{eq:sdot}) are spatially dependent.   Putting $j^A = n^A v$, with $v$ constant in (\ref{eq:var}), one finds vanishing resistivity: $\rho_{xx} = 0$. This is the dissipationless flow of current in a translationally invariant medium.

The velocity $v$ appears quadratically in (\ref{eq:sdot}) and is unconstrained in the variational principle (\ref{eq:var}). Therefore we can `integrate it out' exactly.
This will give a nonlocal functional of the currents $j^A$, due to inverting the viscous terms in (\ref{eq:sdot}). However, it is clear in (\ref{eq:sdot}) that the viscous terms are higher order in derivatives of the velocity field compared to other terms. Therefore, within a strict hydrodynamic expansion in $\ell_\text{ee}/\xi \ll 1$, the viscous terms are subleading and can be dropped. 
We will therefore set $\eta = \zeta = 0$, and minimize over $v$ explicitly.    Using the fact that $\Sigma^{-1}_0$ is symmetric, we obtain
\be\label{eq:JJmin}
\rho_{xx}  = \min_{\nabla \cdot j^A = 0} \frac{\displaystyle  \int \frac{\mathrm{d}^dx}{V} \, j^A R_{AB} j^B }{\left(\displaystyle \int \frac{\mathrm{d}^dx}{V} \, j_x \right)^2} \,,
\ee
where the `incoherent resistivity matrix'
\be\label{eq:Rdef}
R = \Sigma^{-1}_0 P \equiv \Sigma^{-1}_0 \left( 1 - n \,  \frac{n^{\mathrm{T}} \Sigma^{-1}_0}{n^{\mathrm{T}} \Sigma^{-1}_0 n} \right) \,.
\ee
The (skew) projection matrix $P$ satisfies $P n = 0$.    

It is instructive to write $R$ in terms of a matrix of incoherent diffusion constants.   These diffusion constants are found by plugging the linearized constitutive relations (\ref{eq:consgen}) and (\ref{eq:conspi}) into their respective conservation laws.   The resulting hydrodynamic system will have one sound mode, $N-1$ diffusive scalar modes and a diffusive mode for transverse momentum. In Appendix \ref{sec:diffusion} we show that the matrix $R$ appearing in the formula (\ref{eq:JJmin}) can be written in terms of the matrix of
inverse incoherent diffusion constants, $D^{-1}$, as
\be\label{eq:RD}
R = D^{-1}\chi^{-1}.
\ee
One eigenvalue of $D^{-1}$ is zero, and this is associated with the sound mode.  Using (\ref{eq:RD}) in (\ref{eq:JJmin}), the electrical resistivity of a hydrodynamic electron fluid is bounded from above by the incoherent diffusivity matrix $D^{-1}$, which is associated with processes that decouple from momentum drag, cf. \cite{Davison:2015taa}, weighted by a matrix of thermodynamic susceptibilities $\chi^{-1}$.

Our expression (\ref{eq:JJmin}) is exact in the $\xi \rightarrow \infty$ limit, where viscous effects are negligible.  A resistivity bound can be obtained by a variational ansatz in which all the $j_x^A = J_x^A$ are spatially uniform and the $j_y^A = 0$. The constant $J_x^A$ can be taken outside the integrals in (\ref{eq:JJmin}), leading to
\be\label{eq:RR}
\rho_{xx} \leq \min_{J_x^A} \frac{J_x^A \mathsf{R}_{AB} J_x^B}{J_{x}^2} \,, \qquad \mathsf{R}_{AB} \equiv \frac{1}{V} \int \mathrm{d}^dx \, R_{AB} \,.
\ee
Here $\mathsf{R}$ is the spatial average of the incoherent resistivity matrix $R$. The minimization over $J_x^A$ in the above equation can now be performed
exactly, with a simple argument given in Appendix \ref{sec:proof}.  The result is 
\be\label{eq:rhobound}
\rho_{xx} \leq \frac{1}{(\mathsf{R}^{-1})^{nn}} \,.
\ee
Here $(\mathsf{R}^{-1})^{nn}$ is the charge density component of the matrix $\mathsf{R}^{-1}$. This bound is our main result.   In one spatial dimension ($d=1$), where divergence-free currents are necessarily constant, (\ref{eq:rhobound}) is the exact resistivity.
It is possible that with only one incoherent diffusive mode, (\ref{eq:rhobound}) is far too weak, as currents may flow mostly along a percolating set of contours where the ratio of the two densities $n^A$ is held fixed \cite{KS11}.   With multiple incoherent diffusive modes, we expect that (\ref{eq:rhobound}) is tight, up to constant prefactors.

The appearance of diffusivities in the bound can be understood as follows. The background inhomogeneities in the densities $n^A$ mean that a purely advective stationary flow $j^A = n^A v$ is not compatible with simultaneously satisfying all of the conservation laws $\nabla \cdot j^A = 0$. The currents are forced to take the more general form $j^A = n^A v - \Sigma_0^{AB} \nabla \mu_B$ of (\ref{eq:consgen}). The gradients in the $\delta \mu_B$ (and in the conjugate densities $\delta n^A$) sourced in this way then necessarily trigger diffusive dynamics, with an associated entropy production. In addition, to satisfy the conservation laws one must balance the density gradients with the velocity field, which generically requires $\nabla \mu_B \sim \xi^0$. That is to say, the magnitude of the gradients created by the inhomogeneities is independent the wavelength of the inhomogeneities. That is why the inhomogeneity length scale $\xi$ does not appear explicitly in the bound. 

While $R$ is not invertible -- from the definition in (\ref{eq:Rdef}), $R n = 0$ -- the spatially averaged matrix $\mathsf{R}$ appearing in (\ref{eq:rhobound}) is invertible because the null eigenvector of $R$ is rotated from point to point in space. However, in the translationally invariant case, $\mathsf{R} = R$ is not invertible. Thus the bound (\ref{eq:rhobound}) is sufficiently powerful to capture the vanishing resistivity in the absence of momentum relaxation. 

Furthermore, at leading nontrivial order in weak inhomogeneities, the currents minimizing (\ref{eq:JJmin}) will be uniform and hence the bound (\ref{eq:rhobound}) is saturated at leading nontrivial order in this limit.  Focusing on the case where the inhomogeneity is due to fluctuations in the local chemical potential (often called charge puddles in the experimental literature), we show in Appendix \ref{sec:perturbative} that 
\be\label{eq:pert}
\rho_{xx} \approx \frac{\bar V^2}{d \, n^2} \Big(\chi R \chi \Big)_{nn} \,,
\ee
where all quantities can be evaluated in the clean, homogeneous theory, and with the (weak) averaged disorder strength $\bar V^2 \equiv \int \frac{\mathrm{d}^dk}{(2\pi)^d} |V(k)|^2 \,.$ Here $V(k)$ is the Fourier transform of the inhomogeneous chemical potential $\mu(x)$.   The perturbative expression (\ref{eq:pert}) can be obtained by expanding (\ref{eq:rhobound}) or alternatively from a memory matrix computation.    It is straightforward to generalize (\ref{eq:pert}) to alternative types of disorder.

\subsection{Thermal diffusion, phonons and bad metallic SrTiO$_3$}
\label{sec:thermal}

An example clarifies the formal expressions above.   We consider conventional hydrodynamics, where charge, heat and momentum are the only conserved currents (up to disorder).  The one incoherent diffusion mode is thermal.  Previous work has considered the role of thermal diffusion \cite{KS11, PhysRevB.93.075426, Davison:2015taa} in transport.  For the case of thermal diffusion the matrix $R$ takes the form (see Appendix \ref{sec:heat})
\be\label{eq:Rmodel}
R = \frac{T}{\k_0} \left(
\begin{array}{cc}
s^2/n^2 & -s/n \\
-s/n & 1
\end{array}
\right)\,,
\ee
where $\k_0$ is the thermal conductivity. In particular,
the perturbative resistivity (\ref{eq:pert}) is given by (see Appendix \ref{sec:heat} for the susceptibilities)
\be\label{eq:rhothermal}
\rho_{xx} = \frac{\bar V^2}{d} \frac{T}{\k_0} \left(\frac{\pa s/n}{\pa \mu} \right)^2 \,.
\ee
This is precisely the expression obtained in \cite{KS11}. The resistivity (\ref{eq:rhothermal}) is proportional to the inverse of the thermal diffusivity, see again Appendix \ref{sec:heat}. We can now go further and estimate the nonperturbative bound (\ref{eq:rhobound}). A simple assumption is that the inhomogeneities are such that $s,n,\k_0$ in (\ref{eq:Rmodel}) have order one variation over space but each retain the same order of magnitude. In that case, the averaged matrix $\mathsf{R}$ in (\ref{eq:RR}) will have the same form as $R$ in (\ref{eq:Rmodel}), but with order one coefficients appearing in each entry. These coefficients generically render $\mathsf{R}$ invertible and hence from (\ref{eq:rhobound})
\be\label{eq:thermalbound}
\rho_{xx} \lesssim \frac{s^2}{n^2} \frac{T}{\k_0} \,.
\ee
Here $s,n,\k_0$ refer to, for example, their average values. 

For a degenerate Fermi liquid, the temperature scaling of the various quantities in (\ref{eq:thermalbound}) is $s \sim T$ and $\kappa_0 \sim c v_{\mathrm{F}} \tau_{\mathrm{ee}} \sim 1/T$ (here $c\sim T$ is the specific heat),  leading to $\rho_{xx} \lesssim T^4$. At low temperatures this small resistivity will be overwhelmed by the viscous contribution that we dropped in the paragraph above equation (\ref{eq:JJmin}).  While viscous effects are suppressed by the long wavelength of the inhomogeneities, thermal transport is inefficient at low temperatures due to the factors of the entropy density in (\ref{eq:thermalbound});  the $T \to 0$ and $\xi \to \infty$ limits do not commute \cite{us}.   The first signature of hydrodynamic transport at low temperature will be viscous in these cases;  such effects have been observed experimentally  \cite{Molenkamp95,Bandurin1055,Moll1061} and have also been emphasized in theoretical discussions such as
\cite{Gurzhi63,Gurzhi68, KS06, Davison:2013txa, polini, levitov1, levitov2}. At low temperatures, therefore, the transport bound (\ref{eq:thermalbound}) is not accurate, as there will be a further viscous contribution on the right hand side that does not vanish at low temperatures. It is possible that a bound can be found in this regime by balancing viscous and thermal diffusive effects, along the lines suggested in \cite{KS11} (see also discussion below (\ref{eq:rhobound})). The bound (\ref{eq:thermalbound}) is more likely to be relevant at higher temperatures, as we now discuss.

An intriguing possibility allowed by (\ref{eq:thermalbound}) is that non-electronic contributions to the thermal conductivity $\kappa_0$ and entropy $s$ directly influence the electrical resistivity. In particular, the thermal conductivity and entropy can have a significant contribution from phonons. Recent transport experiments on the bad metal regimes of underdoped YBCO \cite{Zhang:2016ofh} and niobium doped SrTiO$_3$ \cite{kamran1,kamran2} have found evidence that $\rho$ is related to the phonon velocity.  The relevance of (\ref{eq:thermalbound}) is especially compelling in the nondegenerate regime of SrTiO$_3$.  At high temperatures the thermal diffusivity of SrTiO$_3$ (with or without a small amount of niobium doping) is measured to go roughly like $D_\text{th.} \sim \hbar v_{\mathrm{s}}^2/(k_{\mathrm{B}} T)$ \cite{kamran2}. Here $v_{\mathrm{s}}$ is the sound speed. In Appendix \ref{sec:heat}, we estimate the temperature scaling $\kappa_0 \sim s D_\text{th.}$. If the bound  (\ref{eq:thermalbound}) is approximately saturated, we obtain 
$\rho_{xx} \sim T \kappa_0/D_\text{th}^2 \sim \kappa_0 T^3 \,.$ The resistivity and thermal conductivity of (doped and undoped) SrTiO$_3$ have been reported in \cite{kamran1, kamran2}: roughly $\kappa \sim T^{-x}$ together with $\rho\sim T^{3-x}$.   At intermediate temperatures (around $100$ K) $x \approx 0$ while at higher temperatures $0 < x < 1$.   This scaling is consistent with our hydrodynamic prediction.

\section{Resistivity from microscopic scattering}

\subsection{Imbalance: Consequences of a non-thermal diffusive mode}
\label{sec:consequences}

\begin{figure}[h]
\centering
\includegraphics[width=6.5in]{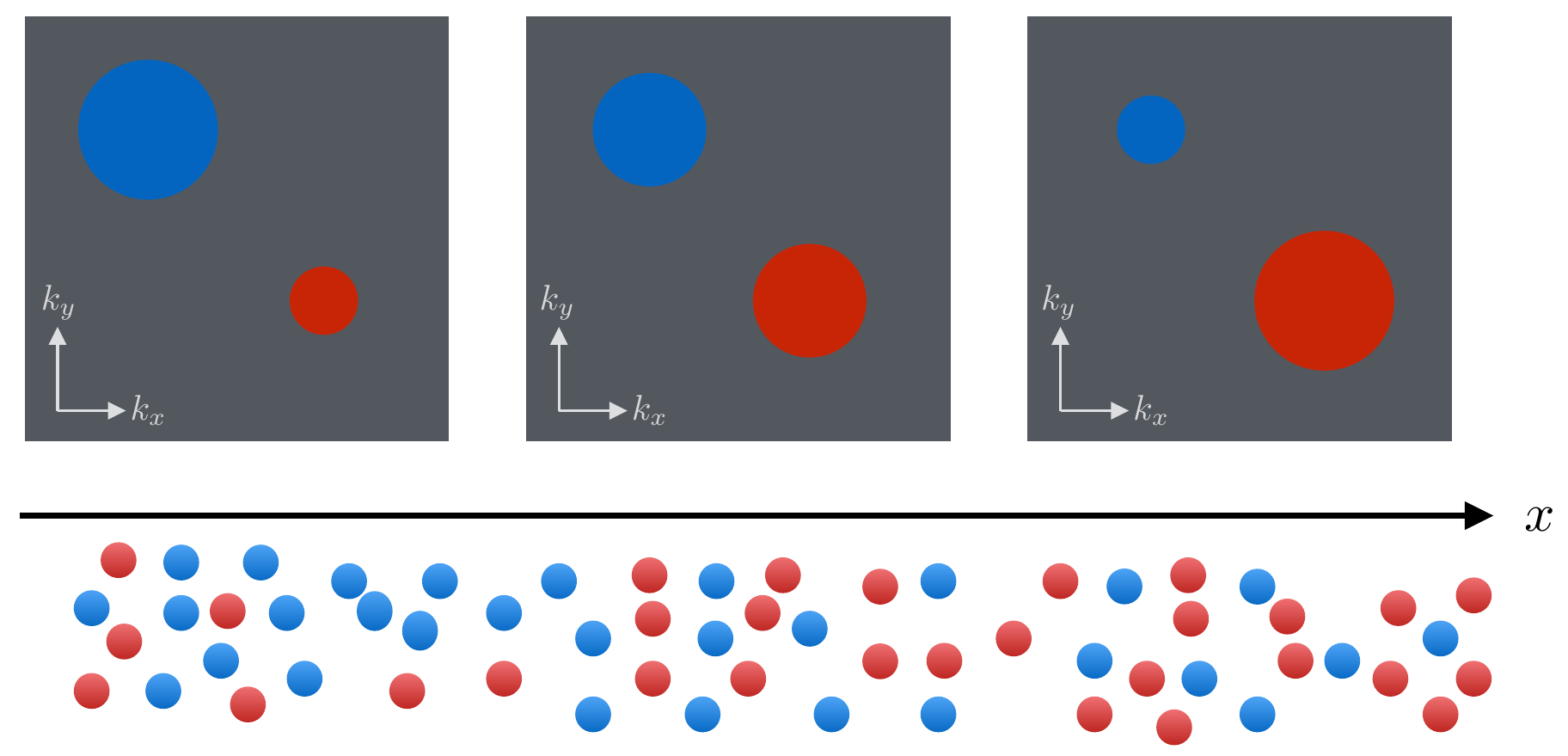}
\caption{Example of a spatial imbalance gradient caused by the inhomogeneous potential.   Top: the Brillouin zone is modified by the disorder, leading to an excess of charge in one pocket relative to another.  Bottom: charge in different pockets clusters in different regions of the disorder potential.   The diffusion of these separately conserved charges through the disorder potential limits transport.}
\label{fig:Imb}
\end{figure}

The transport bound (\ref{eq:rhobound}) is especially powerful in the presence of an additional, non-thermal, charge-carrying diffusive mode.
The contribution of such modes to the resistivity bound will not be suppressed at low temperature by the temperature dependence of the entropy density, as in  (\ref{eq:thermalbound}).  It also becomes more difficult to find parametric improvements to the simple bound (\ref{eq:rhobound}).  An example of a non-thermal diffusive mode is an imbalance mode arising due to the approximate independent conservation of the number of fermions in different bands or pockets: see figure \ref{fig:Imb}.

The key feature of non-thermal electronic diffusive modes is that the corresponding eigenvalues of $\chi$ will approach a finite value as $T\rightarrow 0$. Thus, the temperature dependence of our bound (\ref{eq:rhobound}) is
\be\label{eq:nonthermalbound}
\rho_{xx} \lesssim \frac{1}{\chi_\text{eff} \, D} \,,
\ee
where $D$ is the dominant non-thermal diffusivity in (\ref{eq:RD}) and $\chi_\text{eff}$ is a disorder-dependent constant, proportional to $\chi$, which is temperature-independent at degenerate temperatures.  If (\ref{eq:nonthermalbound}) is approximately saturated, we have achieved the objective of directly relating the resistivity to intrinsic momentum-conserving dynamics.  As emphasized previously, the inhomogeneity length scale $\xi$ does not appear in the resistivity bound (\ref{eq:rhobound}). Instead the resistivity depends on the (temperature dependent) microscopic mean free path $\ell_\text{ee}$. The mean free path determines the diffusivity via $D \sim v_\text{mic} \ell_\text{ee} \sim v_\text{mic}^2 \tau_\text{ee}$. Here $v_\text{mic}$ is a velocity scale of the microscopic degrees of freedom, and $\tau_{\text{ee}}$ is the time scale of momentum-conserving collisions.  We will take $v_\text{mic} \sim v_{\mathrm{F}}$ to be temperature-independent; this can be justified in weakly interacting microscopic approaches such as kinetic theory \cite{us}. Furthermore, angle-resolved photoemission data on bad metals indeed shows that despite significant broadening in energy, sharp features in momentum space survive, with a dispersion about the Fermi surface $\omega \sim v_{\mathrm{F}} |k - k_{\mathrm{F}}|$  \cite{RevModPhys.75.473}.\footnote{
In holographic models of compressible matter, characteristic velocities are instead found to scale as $v_\text{mic} \sim T^{1-1/z}$, with $z$ a dynamic critical exponent \cite{Blake:2016wvh,Roberts:2016wdl}. From the perspective of more conventional electronic systems, such behavior would be indicative of a non-degenerate regime where a single-particle dispersion relation $\omega \sim k^z$ leads to a temperature-dependent velocity $v_\text{mic} \sim \mathrm{d}\omega/\mathrm{d}k \sim k^{z-1} \sim T^{1-1/z}$. This is consistent with the fact that the charge dynamics in holographic models does not show conventional signatures of a degenerate Fermi surface \cite{Hartnoll:2016apf}.}  A further temperature-independent microscopic velocity that can appear is the phonon velocity $v_{\mathrm{s}}$, when electron-phonon coupling is strong.

We emphasize that (\ref{eq:nonthermalbound}) is not a standard Einstein relation in which conductivity and diffusivity are related by $\sigma = \chi D_\text{charge}$. In (\ref{eq:nonthermalbound}), $D$ is \emph{not} the charge diffusion constant, but rather an
incoherent diffusion constant describing processes which are decoupled from the dynamics of momentum.

In the presence of generic disorder, we expect the bound (\ref{eq:nonthermalbound}) will be qualitatively saturated in the presence of imbalance modes. Putting the above formulae together then gives
\be\label{eq:rhosat}
\rho_{xx}(T) \sim \frac{1}{\chi_\text{eff} \, v_{\mathrm{F}}^2} \frac{1}{\tau_\text{ee}(T)} \,, 
\ee
where have made explicit the fact that the temperature dependence of the resistivity is completely inherited from the temperature dependence of the momentum-preserving microscopic scattering rate.

\subsection{From $T^2$ to $T$-linear resistivity}

In a Fermi liquid $\tau_{\mathrm{ee}}(T) \sim \hbar E_{\mathrm{F}}/(k_{\mathrm{B}} T)^2$. Thus (\ref{eq:rhosat}) leads to a novel mechanism for $\rho_{xx} \sim T^2$, without umklapp scattering. In the case where the extra conserved mode is associated with conservation of fermion number in different Fermi pockets or bands, this mechanism is perhaps reminiscent of Baber scattering \cite{Baber383}, where a `light' band which carries the current scatters off a `heavy' band which efficiently relaxes momentum (it is {\it not} sufficient to simply have bands with carriers of different mass).  In contrast, in our hydrodynamic mechanism, the two bands can even have identical masses.  Furthermore, the resistivity is given by (\ref{eq:rhosat}) due to the large imbalance gradients which arise; unlike in Baber scattering, \emph{all} microscopic collisions conserve momentum.

Strong electronic correlations are necessary in order for a Fermi liquid to experimentally exhibit a $T^2$ resistivity. In several classes of compounds that do show a $T^2$ resistivity regime -- including transition metal oxides, heavy fermions and organic metals -- it has been noted that the coefficient of the $T^2$ resistivity can be expressed in terms of electronic properties of the compound \cite{KADOWAKI1986507, Jacko2009}. In particular, this means that the coefficient cannot not have a strong dependence of the amount of disorder in the material. This is compatible with the fact that (\ref{eq:nonthermalbound}) has no singular dependence on the wavelength $\xi$ of the inhomogeneities.   Indeed, many of these materials have complicated band structures with multiple pockets and bands which may provide imbalance modes.   Many of these materials are also layered: dopant impurities are geometrically separated from the conduction layers. This will tend to increase the wavelength $\xi$ of the Coulomb impurity potential created by the impurities \cite{us}.

Furthermore, the phase diagrams of strongly correlated materials showing $T^2$ resistivity are replete with quantum phase transitions and associated `critical fans' exhibiting $T$-linear resistivity. For instance transition metal oxides \cite{husseyA, Hussey, Grigera329}, organic metals \cite{dressel}, heavy fermion compounds \cite{Gegenwart2008,Si1161} and pnictides \cite{pnictide, Analytis2014} show this behavior. For an overview see \cite{Sachdev:2011cs}. Just like the $T^2$ resistivity, the coefficient of the $T$-linear resistivity is known to often be determined by purely electronic properties \cite{andy}, and does not appear sensitive to the purity of the sample.  

As noted previously, quantum critical dynamics is characterized by a `Planckian' \cite{Zaanen2004} microscopic scattering rate $\tau_{\mathrm{ee}}(T) \sim \hbar/(k_{\mathrm{B}} T)$ \cite{subir}.   Using (\ref{eq:rhosat}), we therefore predict a transition from $T^2$ to $T$-linear resistivity upon entering quantum critical fans, due to the change in microscopic scattering rate. The mechanism for momentum relaxation (long wavelength inhomogeneities) does not change. This is an attractive scenario because while the $T^2$ resistivity in these materials could be due to umklapp scattering, there is no compelling candidate for a $T$-linear momentum-relaxing microscopic scattering process at low temperature.

Many $T$-linear materials are in fact bad metals at higher temperatures, with a resistivity above the Mott-Ioffe-Regel limit \cite{PhysRevLett.74.3253,RevModPhys.75.1085, MIR}. Our bound (\ref{eq:nonthermalbound}) is consistent with this behavior, and makes precise the idea that bad metals with $T$-linear resistivity saturate a transport bound \cite{Hartnoll:2014lpa}.  What is novel about our mechanism is that this bound can be saturated without each collision relaxing momentum.    
The diffusion constant for charge in the inhomogeneous theory can be reliably estimated from the incoherent diffusion constants of the clean, homogeneous theory.  Unlike in \cite{Hartnoll:2014lpa}, we do not need to postulate rapid local momentum relaxation.  There is some theoretical \cite{Kovtun:2004de, Maldacena:2015waa} and experimental \cite{Zhang:2016ofh, luciuk2016} evidence that the momentum-conserving scattering rates of clean systems could be bounded $\tau_{\mathrm{ee}} \gtrsim \hbar/(k_{\mathrm{B}} T)$.   Our formalism shows how a Planckian bound on $\tau_{\mathrm{ee}}$ could be directly responsible for a transport bound $\rho \lesssim 1/(\chi_{\mathrm{eff}} D_{\mathrm{inc}}) \lesssim T$.


\section{Outlook}

Without assuming a weakly interacting quasiparticle description, we have described a novel mechanism through which the microscopic scattering rate $\tau^{-1}_{\mathrm{ee}}$ directly controls the temperature dependence of the resistivity.  To determine whether this mechanism is relevant for a specific material, one must go beyond hydrodynamics.  In a complementary paper we show how the bound (\ref{eq:rhobound}) emerges from kinetic theory \cite{us}, in the weak coupling limit.   Within kinetic theory, we can capture both the hydrodynamic regime, as well as the crossover to a ballistic regime when $\xi \ll \ell_{\mathrm{ee}}$.
 
Experimentally, it is crucial to establish whether or not the electron fluids in strange metals are in the hydrodynamic regime.  The objective is to establish the presence of a velocity field in the long wavelength dynamics. This can be more difficult in a non-Fermi liquid with a non-Galilean hydrodynamic limit \cite{Crossno1058}.  If imbalance modes are generic in the hydrodynamic regime of strange metals, they will complicate the search for other signatures of hydrodynamic flow, such as viscous ``whirlpools" \cite{polini, levitov1}, in all but the simplest metals, such as graphene.   

Our proposal motivates searching for evidence of imbalance modes in strange metal transport. This could potentially be achieved by modifying the band structure of the materials to eliminate Fermi pockets, or by direct observation of imbalance diffusion. Previous measurements of diffusive modes in unconventional metals include \cite{Weber2005,Gedik1410, Zhang:2016ofh}.  

It will also be important to identify other non-thermal diffusive modes that can play the required role, including spin imbalance or phase-fluctuating order parameters such as superconductivity \cite{Davison:2016hno} and density waves \cite{Delacretaz:2016ivq}. For instance, the low temperature resistivity of SrTiO$_3$ shows a $T^2$ scaling even when the Fermi surface is too small for umklapp scattering to be effective \cite{kamran}. This system may be an excellent candidate for our non-umklapp mechanism of $T^2$ resistivity, but it remains to identify a candidate non-thermal diffusive mode in this single-band material.

\addcontentsline{toc}{section}{Acknowledgements}
\section*{Acknowledgements}
It is a pleasure to thank Kamran Behnia and Steven Kivelson for helpful discussions. AL was supported by the Gordon and Betty Moore Foundation. SAH is partially supported by a DOE Early Career Award.

\begin{appendix}

\section{Proof of the variational principle}
\label{sec:proof}

There are four statements to prove. \\

\noindent {\bf 1.} The functional (\ref{eq:functional}) in the main text is extremized on stationary solutions to the hydrodynamic equations of motion in the presence of an applied electric field.

\noindent {\bf Proof:} We must extremize (\ref{eq:functional}) with respect to $v$ and $j^A$ and subject to the constraints that $\nabla \cdot j^A = 0$. The constraints are implemented as usual by Lagrange multipliers $\phi_A$. Furthermore, in the presence of `electric fields' $E_A$, the constitutive relation (\ref{eq:consgen})  is modified to
\be\label{eq:mod1}
j^A = n v^A - \Sigma_0^{AB} (\nabla \mu_B - E_B) \,,
\ee
and momentum conservation is modified to
\be\label{eq:mod2}
\dot \pi_i + \pa_j \tau_{ji} = n^A E_A \,.
\ee
We will be interested in only having an electric field sourcing the actual electric current $j_x$, but it is convenient to work with the more general case to start with.

Variation with respect to $v$ directly gives the condition
\be\label{eq:aa}
\pa_j \left(- \zeta \delta_{ij} \nabla \cdot v - \eta \left(\pa_i v_j + \pa_j v_i - \frac{2}{d} \delta_{ij} \nabla \cdot v \right)  \right) = \left(j^A - n^A v\right) (\Sigma_0^{-1})_{AB} n^B \,.
\ee
We can now see that this relation is identically true on solutions to the hydrodynamic equations of motion. Using (\ref{eq:conspi}) in the main text, the term in brackets on the left hand side of (\ref{eq:aa}) is equal to $\tau_{ij} - \delta_{ij} p$. Similarly the term in brackets on the right hand side of (\ref{eq:aa}) is equal to $- \Sigma_0^{AC} (\nabla \mu_C - E_C)$ using (\ref{eq:mod1}). Using (\ref{eq:mod2}) on stationary solutions to set $\pa_j \tau_{ij} = n^A E_A$, along with the thermodynamic identity (\ref{eq:dP}), we see that (\ref{eq:aa}) is simply the generalized Navier-Stokes equation. 

Now we vary with respect to $j^A$. First we consider the currents $j^A$ excluding the electric current:
\be\label{eq:jvar}
(j^A - n^A v) (\Sigma_0^{-1})_{AB} = \left( \int \mathrm{d}^dx \, j_x \right)^2 \nabla \phi_B \,.
\ee
At this point we set $E_A$ sourcing all of these currents (which do not include $j_x$) to zero. Therefore using the constitutive relation (\ref{eq:consgen}) in the main text, the left hand side of (\ref{eq:jvar}) is simply $- \nabla \mu_A$. The equation is therefore solved by setting the Lagrange multipliers
\be
\phi_A = - \frac{\mu_A}{\displaystyle \left( \int \mathrm{d}^dx \, j_x \right)^2} \,.
\ee
For the electric current $j$, the variation leads instead to
\be
(j^A - n^A v) (\Sigma_0^{-1})_{An} = \left( \int \mathrm{d}^dx \, j_x \right)^2 \nabla \phi_n + K \,,
\ee
where the constant $K = [\int \mathrm{d}^dx \left(j^A - n^A v\right) (\Sigma_0^{-1})_{AB}\left(j^B - n^B v\right)]/[\int \mathrm{d}^dx \, j_x]$. Using the modified constitutive relation (\ref{eq:mod1}) for $j$, with an electric field $E_x$, and setting the Lagrange multiplier $\phi_n = - \mu/[\int \mathrm{d}^dx \, j_x]$, the equation is seen to be solved, with the electric field set to be $E_x = K$. \qed\\

\noindent {\bf 2.} The solution to the hydrodynamic equations minimizes the functional.

\noindent {\bf Proof:} The functional (\ref{eq:functional}) in the main text has the abstract form
\be
\mathcal{R}[X] = \frac{X \cdot A \cdot X}{(a \cdot X)^2} \,,  \label{eq:RX}
\ee
for some positive matrix $A$, some vector $a$ and with $X$ a vector. The positivity of $A$ is required in hydrodynamics for positivity of entropy production. The second order variation of $\mathcal{R}[X]$ about a stationary point $X = X_0 + \d X$ is found to be
\be
\rho^{(2)}_{xx} = \frac{\d X \cdot A \cdot \d X}{(a \cdot X_0)^2} + \frac{(a \cdot \d X)^2}{(a \cdot X_0)^3} \,,
\ee
which is manifestly positive for any variation.\qed \\

\noindent {\bf 3.} The variational functional (\ref{eq:functional}) in the main text evaluated on the minimum is equal to the electrical resistivity.

\noindent {\bf Proof:} The variational functional becomes the entropy production evaluated on the solution, divided by the total electric current squared. In the proof of the first statement above we have seen than a uniform electric field is turned on, but there are no other external sources. The entropy production is therefore given by Joule heating: $T \dot s = \sigma_{xx} E_x^2$. Furthermore, in the absence of any external sources except an electric field, the electric current is given by $j_x = \sigma_{xx} E_x$. Evaluating the functional on the solution thus leads to
\be
\rho^{(0)}_{xx} = \frac{\sigma_{xx} E_x^2}{\sigma_{xx}^2 E_x^2} = \frac{1}{\sigma_{xx}} \equiv \rho_{xx} \,,
\ee
as required. \qed \\

\noindent {\bf 4.} The solution to the optimization problem in (\ref{eq:RR}) in the main text is equation (\ref{eq:rhobound}) in the main text.

\noindent {\bf Proof:}  We need to find the minimum value of a functional of the abstract form (\ref{eq:RX}).   A straightforward derivative gives an extremum when \begin{equation}
    \frac{\partial \mathcal{R}}{\partial X} = 0 = \frac{2A\cdot X}{(a\cdot X)^2} - 2a \frac{X\cdot A \cdot X}{(a\cdot X)^3},
\end{equation}
which implies that \begin{equation}
    X = c A^{-1}a.
\end{equation}
with $c$ an arbitrary non-zero constant (since $\mathcal{R}[cX] = \mathcal{R}[X]$).  Hence we find that \begin{equation}
    \min \mathcal{R} = \frac{1}{a\cdot A^{-1} \cdot a},
\end{equation}
which directly leads to (\ref{eq:rhobound}), upon replacing $A$ with $\mathsf{R}$ and $a$ with a unit vector in the $n$ (charge) direction. \qed

\section{Diffusive modes}

\label{sec:diffusion}

As noted in the main text, the hydrodynamic equations of motion lead to
$N-1$ diffusive scalar modes. These are of particular interest, so let us exhibit them explicitly. To isolate the diffusive modes,
 we look for linearized solutions of (\ref{eq:consgen}) and (\ref{eq:conspi}) with $\delta \vec v = \delta P = 0$.   From (\ref{eq:dP}), this means that \begin{equation}
     n^{\mathrm{T}} \mdelta \mu = 0.  \label{eq:ndeltamu}
 \end{equation}
Left-multiplying (\ref{eq:consgen}) by the skew projection matrix $P$ defined in (\ref{eq:Rdef}), and using $\dot{n} = \chi \dot{\mu}$, we arrive at \begin{equation}
    P\chi \mdelta \dot\mu = P\Sigma_0 \nabla^2\mdelta \mu = \left(\Sigma_0 - \frac{n \, n^{\mathrm{T}}}{n^{\mathrm{T}}\Sigma_0^{-1}n}\right) \nabla^2\mdelta\mu = \Sigma_0 P^{\mathrm{T}} \nabla^2\mdelta\mu. 
\end{equation}
Left-multiplying by $\Sigma_0^{-1}$ we obtain (recall that $R$ was defined in (\ref{eq:Rdef}))
\be\label{eq:diff}
R \, \mdelta\dot{n} = R \chi \, \mdelta \dot \mu \equiv D^{-1}\mdelta\dot\mu = P^{\mathrm{T}} \nabla^2 \mdelta \mu  \,.
\ee
The matrix $D^{-1}$ obeys $n^{\mathrm{T}} D^{-1} = 0$.   Hence, one can see that the non-zero eigenvalues of $D^{-1}$ correspond to the incoherent inverse diffusion constants, with (\ref{eq:ndeltamu}) obeyed.

\section{Resistivity at weak disorder}
\label{sec:perturbative}

When momentum relaxation is weak, then the resistivity may be calculated using the memory matrix formalism, by considering the inhomogeneity as a perturbation of the translationally invariant hydrodynamic state \cite{Hartnoll:2016apf}.  The resistivity is given by the Drude-like formula
\be\label{eq:rho2}
\rho_{xx} = \frac{{\mathcal M}}{n^2} \Gamma \,,
\ee
where ${\mathcal M}$ was defined in \S \ref{sec:hydroequations} in the main text and generalizes the role played by the quasiparticle mass density in the usual Drude formula. In the presence of a weak inhomogeneous chemical potential, the momentum relaxation rate is given to leading order by
\be\label{eq:Gamma}
\Gamma = \frac{1}{d {\mathcal M}} \int \frac{\mathrm{d}^dk}{(2\pi)^d} |V(k)|^2 k^2 \lim_{\omega \to 0} \frac{\text{Im} \, G^{\mathrm{R}}_{n n}(\omega,k)}{\omega} \,.
\ee
Here the spectral weight $\text{Im} \, G^{\mathrm{R}}_{nn}(\omega,k)$ is to be calculated in the homogeneous theory. The inhomogeneous source has strength $V(\vec k)$ at wavevector $k$.  In deriving (\ref{eq:Gamma}),  we have assumed that $V(k)$ is isotropic.

When the inhomogeneities are long wavelength (in addition to being weak), so that $\xi/\ell_\text{ee} \gg 1$, then only small wavevectors contribute to the integral in (\ref{eq:Gamma}). In that case, the Green's function $G^{\mathrm{R}}_{nn}(\omega,k)$ that appears can itself be evaluated using (translationally invariant) hydrodynamics  \cite{Kadanoff1963419}. The result can be summarized as follows \cite{Hartnoll:2016apf}. We write the hydrodynamic equations of motion in the form
\be
\dot n^A + M(k)^{AB} \mu_B = 0 \,.
\ee
The matrix of retarded Green's functions for the conserved densities $n^A$ is then
\be
G^{\mathrm{R}}(\omega,k) = M \frac{1}{M - \mathrm{i} \omega \chi} \chi \,.
\ee
Recall that $\chi$ was defined in \S \ref{sec:hydroequations}.  These Green's functions can be explicitly evaluated.  We know that \begin{equation}
    \lim_{\omega\rightarrow0}\frac{\text{Im} \, G^{\mathrm{R}}_{AB}(\omega,k)}{\omega} = \left(\chi M^{-1}\chi\right)_{AB},
\end{equation}
and from (\ref{eq:diff}) that  $M^{-1}_{AB} = k^{-2}R_{AB}$.  Hence,
\be
k^2 \lim_{\omega \to 0} \frac{\text{Im} \, G^{\mathrm{R}}(\omega,k)}{\omega} =  \chi R \chi + \ocal(k^2) \,.
\ee
This result can be substituted into (\ref{eq:Gamma}) and the integral over $k$ performed 
in terms of the averaged strength of inhomogeneities
\be\label{eq:Vbar}
\bar V^2 \equiv \int \frac{\mathrm{d}^dk}{(2\pi)^d} |V(k)|^2 \,.
\ee
Plugging the result for $\Gamma$ into (\ref{eq:rho2}) we obtain (\ref{eq:pert}).

\section{Formulae for thermal diffusion}
\label{sec:heat}

Here we give the derivation of (\ref{eq:Rmodel}) in the main text, and exhibit the connection to thermal diffusion. In minimal linearized hydrodynamics the two conservation laws for the charge and entropy density are
\be
\dot n + \nabla \cdot j = 0 \,, \qquad  \dot s + \nabla \cdot (j^{\mathrm{Q}}/T) = 0  \,.
\ee
Within linearized hydrodynamics, it is convenient to take entropy rather than energy to be the conserved quantity.
The matrix of transport coefficients appearing in the constitutive relations and the matrix of susceptibilities can be written (recall $\pa_\mu s = \pa_T n$)
\be
\Sigma_0 = \left(
\begin{array}{cc}
\s_0 & \alpha_0 \\
\alpha_0 & \bar \k_0/T
\end{array}
\right)\,, \qquad
\chi = \left(
\begin{array}{cc}
\pa_\mu n & \pa_\mu s \\
\pa_\mu s & \pa_T s
\end{array}
\right)\,.
\ee
One then finds from the general formula (\ref{eq:RD}) that the nonzero inverse diffusivity eigenvalue
\be\label{eq:Dthermal}
\frac{1}{D_\text{th.}} =  \left(n \frac{\pa s/n}{\pa T} - s \frac{\pa s/n}{\pa \mu} \right) \frac{T}{\kappa_0} \,,
\ee
where
\be
\k_0 \equiv \bar \k_0 - \frac{2 T s}{n} \a_0 + \frac{T s^2}{n^2} \s_0 \,,
\ee
is the incoherent thermal conductivity with open circuit boundary conditions \cite{Hartnoll:2016apf}. This diffusive mode can therefore be thought of as thermal diffusion. In a Galilean invariant system $\sigma_0 = \alpha_0 = 0$, so that $\k_0 = \bar \k_0$.

From (\ref{eq:Rdef}) in the main text, the matrix
\be
R = \frac{T}{\k_0} \left(
\begin{array}{cc}
s^2/n^2 & -s/n \\
-s/n & 1
\end{array}
\right)\,.
\ee
It is clear that this matrix is proportional to the inverse diffusivity (\ref{eq:Dthermal}).

\end{appendix}

\bibliographystyle{ourbst}
\addcontentsline{toc}{section}{References}

\bibliography{Hydro}

\end{document}